\documentclass[usenatbib,usegraphicx]{mn2e}

\def\Msun{\hbox{$\rm\thinspace M_{\odot}$}}

\usepackage{subfigure}

\title{The Angular Momentum Problem in Cosmological Simulations of Disk Galaxy
Formation}

\author[F.~Piontek \& M.~Steinmetz]{
\parbox{14cm}{Franziska~Piontek and Matthias~Steinmetz}\\
Astrophysikalisches Institut Potsdam, An der Sternwarte 16, 14482 Potsdam, Germany}

\date{}
\pagerange{\pageref{firstpage}--\pageref{lastpage}}
\pubyear{2000}

\begin{document}
\label{firstpage}

\maketitle

\begin{abstract}
We conduct a systematic study of the angular momentum problem in numerical
simulations of disk galaxy formation. We investigate the role of numerical
resolution using a semi-cosmological setup which combines an efficient use of
the number of particles in an isolated halo while preserving the hierarchical build-up
of the disk through the merging of clumps. We perform the same simulation
varying the resolution over 4 orders of magnitude. Independent on the level of
resolution, the loss of angular momentum stays the same and can be tied to
dynamical friction during the build-up phase. This is confirmed in a
cosmological simulation. We also perform simulations including star formation
and star formation and supernova feedback. While the former has no influence
on the angular momentum problem, the latter reduces the loss to a level
potentially in
agreement with observations. This is achieved through a suppression of early
star formation and therefore the formation of a large, slowly rotating
bulge. We conclude that feedback is a critical component to achieve realistic
disk galaxies in cosmological simulations. Numerical resolution is important, but by itself
not capable of solving the angular momentum problem.
\end{abstract}


\section{Introduction}
\label{sec:intro}
The prevailing picture of galaxy formation was formulated almost 30 years ago by \citet{white78} and
\citet{FallEfs1980}. Gas falls into the
potential wells of hierarchically growing dark matter structures and is
additionally governed by dissipational processes. The idea of dark matter
halos acquiring angular momentum through tidal torques from surrounding large scale
structure was first developed by \citet{stromberg} and \citet{hoyle} and later
quantified by \citet{Peebles69}, \citet{Doroshkevich70} and
\citet{white84}. It results in slowly rotating dark matter halos \citep{BE87,MS95}.
The gas within the halo acquires the same specific
angular momentum, but it can cool, therefore collapses and, under of
conservation of angular momentum, forms disks \citep{MMW1998}. Stars are formed due to the
fragmentation of the gas disks, and spheroids are formed during the
interaction and merging of disks \citep{toomre72}. The detailed interaction of
gas, stars and related physical processes make this a very complex problem,
which needs to be addressed through numerical simulations. This is a twofold
challenge. The first challenge is posed by the large dynamic range of scales
which are encountered, from the tens of Mpc scales of halo and galaxy interaction
during assembly to the sub-pc scales of star formation. The second challenge
stems from physical processes like star formation and feedback, which happen on
scales below the resolution of most present day cosmological simulations. It
is thus not so surprising that, despite many efforts in the past years, disk
galaxies with realistic properties remain elusive in simulated universes. Especially bulge-less
disks, which have been shown to exist in non-negligible numbers
\citep{kautsch} in the nearby universe, have not been produced in a
simulation up to this point. \par
One of the major problems encountered in numerical simulations of disk formation is the
so called angular momentum catastrophe \citep{NavarroBenz1991}. Early
simulations all resulted in small, highly centrally concentrated disks with dominating bulges
\citep[for example][]{NavarroWhite1994,MSNavarro1999,NavarroMS2000} after loosing large
amounts of angular momentum during the formation. As four main causes
have been proposed: (1) lack of resolution \citep{fabio2004, Kaufmann2007};
(2) an artificial transfer of angular momentum from the cold disk at the
interface to the hot
halo \citep{okamoto03}; (3) disk instabilities leading to fragmentation
\citep{Robertson2004}; (4) merging of substructure leading to transfer of
angular momentum to the outer halo via dynamical friction  \citep{NavarroBenz1991,NavarroWhite1994}. The second and third effect can be
avoided with an improved implementation of gas physics (a decoupling of hot and cold gas
phases for the second effect as suggested by \citet{okamoto03} or a
multiphase medium for the third effect according to
\citet{Robertson2004}). Low numerical resolution, the first effect, can cause
artificial loss of angular momentum through a variety of effects like numerical
two-body heating due to the large mass difference between dark matter and gas
particles \citep{MSWhite1997,Mayer2005}, artificial viscosity, an 
incorrect pressure estimate in the standard SPH  method for particles with
significant temperature differences (as happens for example at the edge of the
disk), and an artificial
asymmetry of the disk causing gravitational torques from the surrounding
halo. These problems can be remedied with larger numbers of particles in the
disk. Others, like a correct modeling of the disk center where velocity
gradients are steep and angular momentum is easily lost, or the formation of
bars, which can transport angular momentum outward efficiently, are difficult to
address even with very large particle numbers. While these effects have been
demonstrated to operate in well established disks and smooth collapse
scenarios \citep{Kaufmann2007}, it still needs to be demonstrated to what
extend they are responsible for the angular momentum loss seen in fully
cosmological large scale structure simulations.\par
As opposed to the first three effects, the fourth, merging of substructure, is
a physical effect which is
natural to $\Lambda$CDM. It affects dark matter as well as gas, and angular
momentum is lost through dynamical friction of interacting and merging
clumps. The most efficient mechanism to prevent angular momentum loss through
this process is to prevent the formation of too many clumps. For this, two
scenarios have been discussed. The first one involves a change in the
cosmological model from cold to warm dark matter (WDM). While leaving the
large scales untouched, this results in a smoothing of small scales of the
power spectrum, and therefore less substructure and a smoother accretion of
baryons onto dark matter halos instead of dense clumps susceptible to dynamical
friction. \citet{SommerLarsen2001} successfully showed that this leads to an
alleviation of the angular momentum problem, which puts doubt on the claim that
the angular momentum problem is mainly a resolution issue. It can also help to
solve what is known as the "missing satellite problem" in $\Lambda$CDM, a large excess in
dark matter substructure predicted by numerical simulations compared to
observed numbers of satellite galaxies \citep{Goetz2003}. However,
observational constraints on the mass of dark matter particles, for example by
the redshift of re-ionization (which is delayed significantly in a WDM universe
due to the lack of low mass halos), already allow only varieties which are
tepid at most \citep{Viel2008}. In the second scenario preventing
the gas from forming small, dense clumps before entering the main halo, the
thermal energy of the gas is increased via physical processes like feedback from
supernova explosions or stellar winds. We will address this in the second part
of this paper. \par
In this paper, we will focus on the role of resolution and feedback with
respect to the angular momentum problem in numerical simulations of disk
galaxy formation. We present a detailed, systematic study of resolution
effects in isolated but semi-cosmological halos preserving the essence of the
hierarchical formation history, and also test the results in fully
cosmological runs. The study also covers the effects of various physical processes ranging from
simulations including only radiative cooling over the inclusion of simple star
formation to the application of a model for supernova feedback. The paper is organized as follows: in $\S$2 we describe the code and the
initial conditions we use. In $\S$3 we present the results on the angular
momentum problem in simulations including only gravity and radiative
cooling. $\S$4 discusses the impact of star formation and the feedback model
and in $\S$5 we summarize and discuss our results.

\section{Code and initial conditions}
All simulations are based on the WMAP3 cosmology
\citep{WMAP3year} with H$_0$=73 km s$^{-1}$Mpc$^{-1}$, $\sigma_8=0.75$,
n$_{\rm{S}}=0.9$, $\Omega_0=0.24$, $\Omega_{\Lambda}=0.76$ and $\Omega_{\rm{b}}=0.04$. We
use the N-body code \textsc{GADGET2} \citep{Springel2005} in a version
including radiative cooling based on \citet{Katz96}. To minimize the number of
free parameters we do not include a UV
background. The gas is treated within the Smoothed Particle Hydrodynamics (SPH)
framework (\citet{gm77}, \citet{lucy77}). In the following, we
explain our implementation of star formation and supernova feedback.\par
\subsection{Star formation and feedback} 
\label{subsec:code}
Our basic star formation model follows \citet{Katz1992} and \citet{NavarroWhite1994}. The main criteria for either spawning a new star particle from a gas particle or turning the gas particle into a star particle (depending on the gas mass) are:\\
(1) $\rho>\rho_{\rm{crit}}$, where we choose\footnote{This choice of density corresponds to a region where the cooling
time is always shorter than the dynamical time, resulting in rapid cooling to
the cut-off temperature of the cooling curve (10$^4$K) where the Jeans instability criterion (sound crossing time is larger than the
dynamical time) is  almost
always fulfilled \citep{NavarroWhite1994}.} $\rho_{\rm{crit}}$=7$\times10^{-26}$g
cm$^{-3}$\\
(2) $\bigtriangledown \overrightarrow{v}<0$\\
(3) T$<3\times10^{4}$ K.\\
The star formation rate is guided by a basic Schmidt law \citep{schmidt59},
but expressed in physical density as
\begin{equation}
\frac{d\rho_{\star}}{dt} = c_{\star} \frac{\rho_{gas}}{t_{\star}}
\end{equation}
where $c_{\star}$ is a dimensionless parameter regulating the star formation
efficiency (we use $c_{\star}$=0.1, in agreement with observations
by \citet{Duerr1982} for the ISM in the Milky Way) and 
\begin{equation}
t_{\star}=max(t_{dyn},t_{cool}),\,\,t_{dyn}=(4\pi G \rho)^{-1/2}.
\end{equation}
We use a stochastic approach to star formation, so the probability to form a
star particle during a given timestep $\bigtriangleup$t is given by
\begin{equation}
p=\frac{m_{gas}}{m_{gas,orig}\epsilon}(1-e^{-c_{\star}\frac{\Delta t}{t_{\star}}}).
\end{equation}
$\epsilon$ is a parameter giving the fraction of the gas mass turned into a
star or, alternatively interpreted, the number of generations of stars. We generally use
$\epsilon$=2 to prevent too much increase in the number of particles.\par
We include the most basic but also probably most
influential type of stellar feedback: energy output from the late stages of
the evolution of massive stars (M$>$8 $\Msun$). This is usually referred to as
supernova feedback, but actually covers all types of energetic feedback which are
not resolved in this study. The model implemented here showed the best performance in a
more extensive study of feedback models which is presented in
\citet{Paper2}. As in all N-body simulations, each ``star particle''
represents a whole population of stars of total mass $\rm{M}_{\star}$ (which
depends on resolution). For this we assume a Miller-Scalo initial mass
function (IMF) \citep[IMF,][]{millerscalo} with a lower cutoff of 0.1 M$_{\odot}$ and an upper
cutoff of 100 M$_{\odot}$: 
\begin{equation}
\xi(M)= M_{\star}A \left\{\begin{array}{cl} M^{-1.25} & 0.1 < M < 1 \Msun\\
    M^{-2} & 1 < M < 2 \Msun\\ 2^{0.3} M^{-2.3} & 2 < M < 10 \Msun\\ 10 \
    2^{0.3} M^{-3.3} & 10 < M < 100 \Msun \end{array}\right.
\end{equation}
where A=0.284350751 and $M_{\star}$ is the mass of the stellar population, in
our case of the star particle.
Stars with masses between 8 and 100 M$_{\odot}$
explode as type II supernova and each explosion is assumed to yield an energy of 10$^{51}$
ergs. This corresponds to an energy of 1.21$\times10^{49}$ erg per solar mass
formed. To avoid numerical artefacts related to abrupt changes in the thermal
properties of the gas which cannot be resolved, the energy is returned to the surrounding gas over time following an exponential approach using
\begin{equation}
\Delta E(t) = E_{SN} \frac{t-t_{\star}}{t_{SN}} e^{- \frac{t-t_{\star}}{t_{SN}}} \frac{\Delta t}{t_{SN}}.
\end{equation}
Here, $\rm{E_{SN}}$ is the total supernova energy, $\rm{t_{\star}}$ is the time when the
star particle was created and $\rm{t_{SN}}$ is a typical time scale (we use 20
Myr as the lifetime of a 8~$\Msun$ star). This energy is smoothed over the neighboring gas particles using the
SPH smoothing kernel, so each neighbor gets an energy of 
\begin{equation}
\Delta E_{SN,i}=\Delta E(t) \frac{W(| \overrightarrow{r}_{i}-\overrightarrow{r}_{\star}|,h_{\star})M_{i}}{\rho_{\star}}.
\end{equation}
$\rm{h_{\star}}$ is the smoothing length and the most important parameter
here. This is a variable length scale calculated similar to the gas smoothing
length in \textsc{GADGET2} (the mass within the smoothing sphere is kept constant). \par
\citet{Katz1992} and \citet{MSMueller95} found that supernova feedback energy is radiated away
very quickly in the simulations, since star formation by definition occurs in
dense regions with short cooling times. Consequently, in these early
simulations, feedback had little impact. In more realistic ISM models
this does not happen due to the multiphase nature of the gas, but it is
computationally demanding, if at all possible, to achieve enough resolution to directly simulate this.
One method to prevent this immediate re-radiation was suggested by
\citet{Gerritsen1997} and later \citet{thack00}. It involves turning off radiative
cooling in gas particles receiving feedback energy and evolving them
adiabatically for a given time. We adopt this method and turn off the cooling
locally for a time of 20 Myr (in agreement with earlier simulations and
similar to the lifetime of a supernova shell).  
\subsection{Isolated initial conditions}
It is our goal to isolate the root cause of and find a solution for the angular momentum
problem. In a fully cosmological simulation, this is difficult, since the
resolution for a given halo and particularly for the forming galaxy is
limited and the hierarchical merging processes are likely to destroy disk progenitors.
Therefore we first perform a controlled experiment for an
isolated halo, and then verify our conclusions in a cosmological box. However,
a completely isolated and idealized setup consisting of an NFW halo\footnote{A
halo with the density profile as in equation~\ref{eq:NFW}, quantified by \citet{Navarro1997}.}
in combination with a hot gas halo, as in \citet{Kaufmann2007}, is too artificial to probe the
relevant physics, since it completely lacks any merging in the formation
process. To be more realistic with the same benefits of high resolution in the
gas disk and no major mergers destroying it, we use a semi-cosmological setup
as in \citet{Katz1991}. We start with a cosmological box with a side length of
2 h$^{-1}$Mpc, large enough to host one Milky Way-type halo, but lacking
all large scale structure. Centering the forming halo, we cut out a sphere
with 1 h$^{-1}$Mpc radius which is simulated in physical coordinates and with 
vacuum boundary conditions. To mimic the missing large scale effects, we apply
an overdensity of $3\sigma$ and put the sphere into solid body rotation with a
spin parameter of $\lambda=0.08$. This is higher than what is typical
for dark matter halos \citep[$\lambda\approx0.04$][]{bullock01}, but will
ensure the formation of a large rotationally supported and stable gas disk. We
include gas particles by splitting each dark matter particle into a gas and a dark matter particle with the mass ratio given by the density
parameters. The initial redshift of the simulations is z=75. For our
resolution study, we create the same initial conditions with four different
resolutions, based on boxes with 16$^3$, 32$^3$, 64$^3$ and 128$^3$ dark
matter particles. This corresponds to spheres with $2\times10^3$,
$2\times10^4$, $1.4\times10^5$ and $1.1\times10^6$ particles each in gas and
dark matter. We use a gravitational softening growing proportional to the
scale factor for z$>$10 and staying fixed in physical coordinates after z=10, at a value
of 1h$^{-1}$kpc for the 3 lower resolutions and 0.25 h$^{-1}$kpc for the
highest resolution, for both gas and dark matter particles. We also tested
higher softenings for the two lowest resolutions - 4 kpc
h$^{-1}$ for 16$^3$ and 2 kpc h$^{-1}$ for 32$^3$, which is in better
agreement with the criterion derived by \citet{Power2003}. However, the
results for these runs do not differ significantly from those with
smaller softening length, so they are not presented here. 

\subsection{Cosmological initial conditions}
From a cosmological 64$^3$~h$^{-1}$Mpc box we resimulate a halo promising to
host a Milky Way-type galaxy. The crucial selection features are the halo mass
($\rm{M_{halo}}\simeq10^{12}\,\rm{h}^{-1}\,\Msun$), a relatively quiet merging history (the
redshift of the last major merger with mass ratios of 1:3 and 1:10 are z=3.63 and z=2 respectively), a high spin parameter ($\lambda=0.057$) and no
halo with equal or larger mass within a distance of 2 Mpc. We perform runs
with three levels of resolution in the resimulated region corresponding to an
effective resolution of 512$^3$, 1024$^3$ and 2048$^3$. This corresponds to
particle masses for dark matter (gas) of $1.47\times10^8\Msun$
($3.1\times10^7\Msun$), $1.86\times10^7\Msun$
($3.71\times10^6\Msun$) and  $2.32\times10^6\Msun$
($4.64\times10^5\Msun$) respectively. Star particles have half the mass of the
initial gas particle mass. A gravitational softening of 4 (3), 2
(1.5) and 1 (0.75) h$^{-1}$ kpc is used.

\subsection{Analysis of a simulated galaxy}
\label{subsec:amcalc}
%We define our halos using $R_{200}$ which is the radius where
%$\rho(r)=200\rho_c$ with $\rho_c=3H_0^2/(8\pi G)$. As gaseous disk we define
We define our halos using the virial radius $\rm{R_{vir}}$, which is the radius where
$\rho(r)=\Delta\rho_c$ with $\rho_c=3H_0^2/(8\pi G)$ and
$\Delta=18\pi^2+82x-39x^2$, $x=\Omega_0(1+z)^3/(\Omega_0(1+z)^3+\Omega_{\Lambda})-1$.
The cold gas ($T<3\times10^4$K) in the center of the halo makes up a clearly defined disk. In runs without
star formation and feedback, all cold gas ends up in that disk. In the runs
with feedback, some hot gas surrounding the cold gas is present, but the disk can still be distinguished clearly. When star formation is included, stars within the disk radius are
defined as being part of the galaxy. Since the rotationally supported gas disk
gives a clear axis of symmetry, we rotate the system to have this axis aligned
with the axis in z direction and the disk in the x-y plane.
The total specific angular momentum j of the disk is then computed by summing
over all disk particles, with respect to the center of mass of
these particles. It is given by
\begin{equation}
j=\frac{|\overrightarrow{J}|}{\sum m_i}
\end{equation}
where
\begin{equation}
\overrightarrow{J}=\sum m_i (\overrightarrow{r}_i-\overrightarrow{R}_{\rm{COM}})\times(\overrightarrow{v}_i-\overrightarrow{v}_{\rm{COM}}).
\end{equation}
To compute the time evolution of the angular momentum, all particles belonging to
the galaxy at z=0 are tracked back in time and the angular momentum is
calculated in earlier snapshots for these particles with respect to the center
of mass at the time. When star formation is included, the progenitor gas
particles are used at the time when stars have not formed yet.

\section{The angular momentum problem in simulations with radiative cooling}
Our first, most basic set of runs includes dark matter and gas and the latter
is subject to radiative cooling only. Star formation and feedback are turned off. This
should enable us to pinpoint the processes responsible for the loss of angular
momentum and investigate the role resolution can play to avoid
this process without the additional, probably positive effects feedback should have.  
\subsection{Isolated initial conditions}
We first analyze the results of our controlled semi-cosmological, isolated
experiment in four different resolutions. Some characteristic parameters of the
different runs are summarized in Table~\ref{tab:stats}. 
\begin{table*}
\begin{tabular}{rrrrrrrr}
\hline
\hline
&16$^{3}$&32$^{3}$&64$^{3}$&128$^{3}$&cosmological run\\
\hline
\hline
DM~particle~mass~[$M_{\odot}$]&3.62$\times10^{8}$&4.5$\times10^{7}$&5.64$\times10^{6}$&7.05$\times10^{5}$&1.86$\times10^7$\\
\hline
gas~particle~mass~[$M_{\odot}$]&7.24$\times10^{7}$&9.01$\times10^{6}$&1.13$\times10^{6}$&1.41$\times10^{5}$&3.71$\times10^6$\\
\hline
\#~of~gas~particles~in~disk&1510&12710&106319&606895 (z=1.7)&38856\\
\hline
$\epsilon$\,[h$^{-1}$ kpc]&1&1&1&0.25&1.5\\
\hline
R$_{\rm{vir}}$\,[kpc]&248&251&251&&292\\
\hline
V$_{\rm{vir}}$\,[km s$^{-1}$]&123.38&124.45&124.55&&145.45\\
\hline
M$_{\rm{vir}}$\,[M$_{\odot}$]&8.73$\times10^{11}$&8.99$\times10^{11}$&9.01$\times10^{11}$&&1.21$\times10^{12}$\\
\hline
c&28.04&29.68&30.542&&10.13\\
\hline
$\lambda$&0.083&0.088&0.083&&0.055\\
\hline
\hline
\end{tabular}
\caption{Characteristic parameters for the semi-cosmological simulations with
  increasing resolution and for the cosmological run
  with an effective resolution of 1024$^3$ particles in the high resolution
  region as used in the run without star formation and feedback.}
\label{tab:stats}
\end{table*} 

\subsubsection{The resulting disk}
\begin{figure}
\begin{center}
\includegraphics[width=0.5\textwidth]{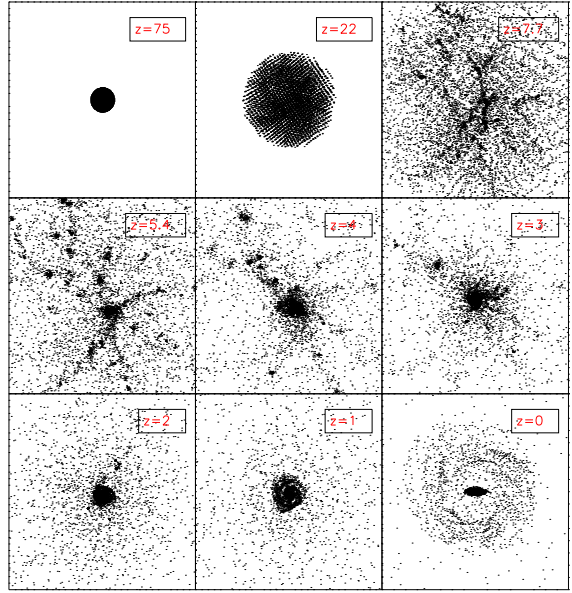}
\caption{The formation of the disk in our 32$^{3}$ sphere from the starting
  redshift of 75 until z=0. Each plot has a side length of 280 physical kpc except
  the last one which zooms in to 80 kpc side length.}
\label{fig:semidiskevol}
\end{center}
\end{figure}

The evolution in the semi-cosmological setup is illustrated in
Figure~\ref{fig:semidiskevol} for the 32$^3$ resolution. Since it
is a physical coordinate system, initially the sphere expands following the
expansion of the universe. The redshifts of turnaround at z$\sim$12 and collapse at z$\sim$8 agree well with
analytical expectations based on \citet{Lokas2001}. Small, cold gas clumps
merge to form a central disk. This process is finished around z=1 after
which there is no longer any significant evolution. There is no late infall owing to
the missing large scale environment. The disk can be described as a central
overdense clump with a radius of about 5 kpc containing up to 80\% of the
total disk mass, and a thin outer disk. The resulting disk is unchanged with increasing resolution. \par
\begin{figure}
\center
\includegraphics[width=0.5\textwidth]{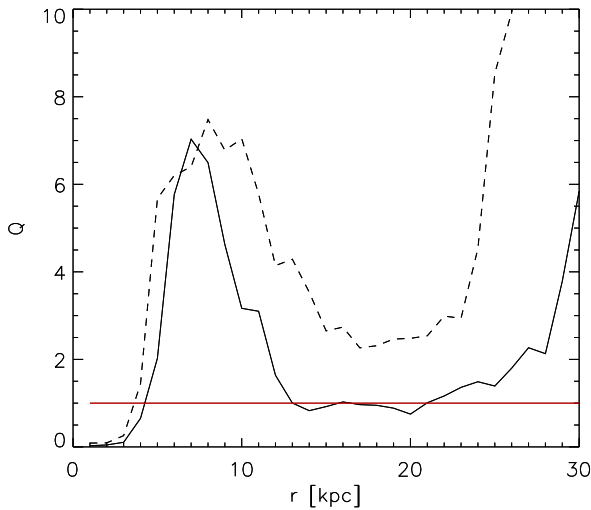}
\caption{Disk stability is measured with the Toomre Q parameter in our 32$^{3}$
  sphere. The solid horizontal line shows the stability criterion $\rm{Q}>1$. The
  solid curve is for the standard model, the dashed curve for the model with
  $5\times10^{4}$ K minimum gas temperature.}
\label{fig:toomre}
\end{figure}

We examine the stability of the disk with respect to the Toomre Q parameter \citep{toomre}
\begin{equation}
Q(r)=\frac{c_{s} \kappa}{\pi G \Sigma_{g}}
\end{equation}
where c$_{s}$ is the effective soundspeed, $\kappa$ the epicyclic frequency
and $\Sigma_{g}$ the gas surface density. The disk is stable where
Q(r)$>$1.0. As illustrated in
Figure~\ref{fig:toomre}, in our case the disk is clearly unstable in the
innermost, compact  region, and only marginally stable in the outer
parts (solid curve). Star formation can therefore happen in the disk. It could be stabilized if the gas were to remain hot with a minimum temperature of $5\times10^{4}$ K, as shown
by the dashed line in Figure~\ref{fig:toomre}.

\subsubsection{Verification of the setup with an analytical model}
\begin{figure}
\begin{center}
\includegraphics[width=0.5\textwidth]{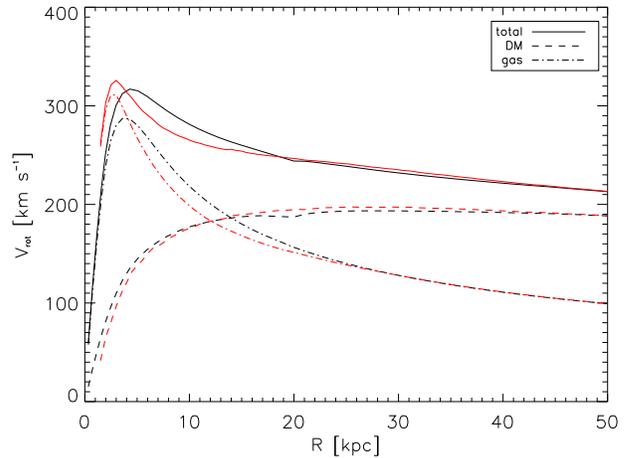}
\caption{Comparison of the rotation curves from the simulation (red) and
  predicted by the MMW model (black).}
\label{fig:MMW}
\end{center}
\end{figure}
We attempt a verification of our semi-cosmological initial conditions setup by comparing the
resulting disk to predictions obtained with the analytical galaxy formation
model by \citet{MMW1998} (hereafter MMW). In their model MMW derive the
properties of disks (like mass, scale radius, central surface density) from
the density profile of the dark matter halo and its angular momentum under the
following four assumptions: (1) the disk mass is a fixed fraction of the halo
mass (m$_{\rm{d}}$=M$_{\rm{disk}}$/M$_{\rm{halo}}$), (2) the disk angular
momentum is a fixed fraction of the halo angular momentum
(j$_{\rm{d}}$=J$_{\rm{disk}}$/J$_{\rm{halo}}$), (3) the disk is thin,
centrifugally supported and has an exponential surface density profile, (4)
only dynamically stable systems can correspond to real galaxies. The disk
properties are then completely determined by the rotational velocity of the
halo at r$_{200}$ (the radius where $\rho(r)=200\rho_c$),
V$_{200}$, the concentration c of the halo, the spin parameter $\lambda$, m$_{\rm{d}}$ and j$_{\rm{d}}$, and
are calculated iteratively taking into account the influence of the forming
disk on the halo density profile and rotation curve. This is a gravitational
influence as well as a contraction of the inner regions of the halo. \\
As input parameters we use the results from our
simulations, the disk mass $\rm{M_{d}}$ and disk angular momentum J$_{\rm{disk}}$ from the run with
radiative cooling, and the dark matter halo parameters (halo mass M$_{200}$, halo angular
momentum J$_{200}$, halo concentration c, spin parameter $\lambda$) from the
pure dark matter run (note the use of R$_{200}$ instead of R$_{\rm{vir}}$
here, to define the halo in agreement with MMW). For our purposes we
mainly look at the predicted rotation curves for the disk and dark matter halo, which we then compare to the
rotation curves obtained in our simulations. Table~\ref{tab:MMWstats} gives
the MMW parameters
for the different runs. We applied two major modifications to the original MMW
model. MMW assume an NFW halo with a density profile given by
\begin{equation}
  \rho(r)=\rho_c\frac{\delta_c}{(r/r_s)(1+r/r_s)^2},
  \label{eq:NFW}
\end{equation}
where $\delta_c$ is a characteristic density and $r_s$ is a scale length.
Since our initial conditions are only semi-cosmological and we are
missing the surrounding large scale structure, we are also missing
some late infall of material into our halo which would happen in a fully 
cosmological box. This results in density profiles which deviate
from a typical NFW profile at large radii. Therefore, instead of
using a halo mass profile based on the NFW profile as 
done by MMW, we use the calculated mass profile of the dark matter halo as an input
for the model.\par
 A second problem we encountered was a systematic over-prediction
of the inner part of the dark matter rotation curve by the MMW model. As
already mentioned above, the mass
profile of the halo is influenced by the formation of the disk. This effect
is described by the adiabatic contraction model, based on work by \citet{BW84}
and \citet{Blum86}. In this model, particles are assumed to move on circular
orbits and angular momentum is strictly conserved. For circular orbits, the
force is $F/m=v^2/r$ and the angular momentum is $j/m=vr$. Combining them and
including angular momentum conservation leads to the general result for a
particle starting at radius r$_{\rm{i}}$ and ending at radius r$_{\rm{f}}$
\begin{equation}
F_{f}(r_{f})r_{f}^3=F(r_{i})r_{i}^3.
\end{equation}
With the Newtonian force law F/m=$\frac{GM(r)}{r^{2}}$, this leads to
$M_f(r_f)r_f=M_i(r_i)r_i$, where M(r) is the enclosed mass at radius r. This
is the form used by MMW. 
However, in order to reduce the
occurrence of two-body direct interactions in the simulations, the forces are softened and may be
described rather with a Plummer force law
\begin{equation}
F_{Plummer}=\frac{-GMmr}{(r^{2}+\epsilon^{2})^{3/2}}.
\end{equation}
We therefore modify the MMW model by using the force profile instead of the
mass profile to calculate the adiabatic contraction \citep{Gottbrath}. The force profile is
given by 
\begin{equation}
F(r)=\int F(S,r)dS
\end{equation}
 where F(S,r) is the Plummer force profile of a spherical shell with radius S
 which in turn is given by
\begin{equation}
\frac{dF}{dS}=-\frac{GmM}{2r^{2}S}\left(\frac{\epsilon^{2}+S(r+S)}{\sqrt{\epsilon^{2}+(r+S)^{2}}}-\frac{\epsilon^{2}+S(S-r)}{\sqrt{\epsilon^{2}+(r-S)^{2}}}\right).
\end{equation}
With this modified approach we achieve good agreement between the rotation
curves predicted from the model and produced in the simulation, as can be seen
in Figure~\ref{fig:MMW}. The only
region being different is the inner 15 kpc of the disk rotation curve, which
is much more peaked in the simulations. This is due to the high concentration
of matter in the center as a result of the angular momentum problem, and to a lack of resolution in the center. The
central clump behaves like a bar, sweeping up gas directly surrounding
it and resulting in an underdense shell between the central clump and the thin
disk. This is also visible in the bottom right plot of the disk in
Figure~\ref{fig:semidiskevol}. Therefore, the rotation curve is too peaked in the very
center, and too low between radii of 5 and 18 kpc.
\begin{table*}
%\begin{tabular}{rrrrrr}
%\hline
%\hline
%&16$^{3}$&32$^{3}$&64$^{3}$\\
%\hline
%\hline
%R$_{200}$\,[kpc]&189&191&192\\
%\hline
%V$_{200}$\,[km s$^{-1}$]&138.24&139.42&139.47\\
%\hline
%M$_{200}$\,[M$_{\odot}$]&8.36$\times10^{11}$&8.59$\times10^{11}$&8.64$\times10^{11}$\\
%\hline
%j$_{200}$\,[kpc km s$^{-1}$]&2233.25&2100.72&2230.24\\
%\hline
%M$_{\rm{disk}}$\,[$M_{\odot}$]&1.09$\times10^{11}$&1.15$\times10^{11}$&1.2$\times10^{11}$\\
%\hline
%j$_{\rm{disk}}$\,[kpc km s$^{-1}$]&1017.6&1092.4&1234.8\\
%\hline
%m$_{\rm{d}}$&0.13&0.134&0.139\\
%\hline
%j$_{\rm{d}}$&0.059&0.07&0.077\\
%\hline
%\hline
%$\lambda$'=(j$_{d}$/m$_{d}$)$\lambda$&0.035&0.037&0.042&0.05&0.045\\
%\hline
\begin{tabular}{rrrrrrrrr}
\hline
\hline
resolution&R$_{200}$&V$_{200}$&M$_{200}$&j$_{200}$&$_{\rm{disk}}$&j$_{\rm{disk}}$&m$_{\rm{d}}$&j$_{\rm{d}}$\\
&[kpc]&[km s$^{-1}$]&[M$_{\odot}$]&[kpc km s$^{-1}$]&[$M_{\odot}$]&[kpc km
s$^{-1}$]&&\\
\hline
\hline
16$^{3}$&189&138.24&8.36$\times10^{11}$&2233.25&1.09$\times10^{11}$&1017.6&0.13&0.059\\
\hline
32$^{3}$&191&139.42&8.59$\times10^{11}$&2100.72&1.15$\times10^{11}$&1092.4&0.134&0.07\\
\hline
64$^{3}$&192&139.47&8.64$\times10^{11}$&2230.24&1.2$\times10^{11}$&1234.8&0.139&0.077\\
\hline
\hline
\end{tabular}
\caption{Parameters for the MMW model.}
\label{tab:MMWstats}
\end{table*} 

\subsubsection{The angular momentum analysis}
\begin{figure}
\begin{center}
\includegraphics[width=0.5\textwidth]{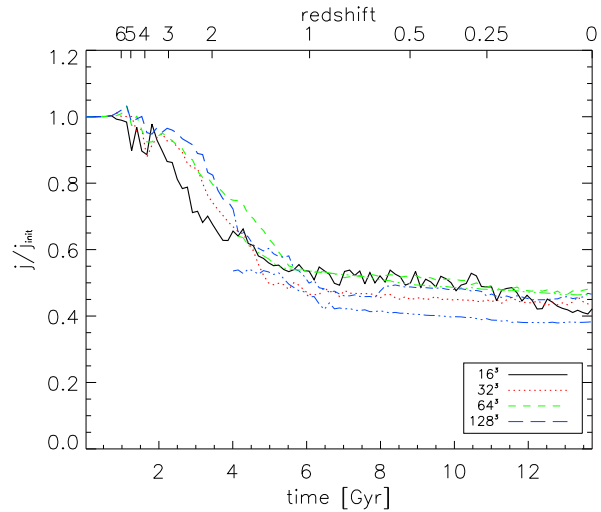}
\caption{The evolution of the specific angular momentum over time for the cold
  gas within the virial radius. The different colored lines show the 4
  different resolutions of the semi-cosmological halos. The dash-dotted and
  dash-double-dotted parts of the blue line indicate the evolution for the
  hightes resolution case, when gas was converted to stars, calculated in two
  different ways. As a comparison, this was also done for the 64$^3$
  resolution, shown in the green dash-double-dotted line.}
\label{fig:AMcomp}
\end{center}
\end{figure}
We analyze the loss of angular momentum by calculating in all snapshots the
specific angular momentum of cold gas particles within a radius of 250 kpc
(which is the virial radius at z=0). Since most cold gas particles end up in
the disk, this basically traces the angular momentum evolution of disk
particles. The result is shown in Figure~\ref{fig:AMcomp}. The highest resolution
run (128$^3$ particles in the underlying box, blue line) poses a computational
challenge. Without star formation, it is computationally infeasible to
advance the simulation beyond z$\sim$1.7. We overcome this by replacing all gas
particles at this point with collisionless ``star'' particles. In this way, no
further cooling is possible, but the disk is already largely in place at this
point. Reducing the computation to only gravitational forces enables us to
extend the run until z=0. For the angular momentum calculation unilt z=1.7, we
use the cold gas within the virial radius, as in the other resolutions. After
that we follow two different methods. First we tested to use all star
particles within the virial radius (thin, dash-dotted blue line). However, in a
comparison of the same procedure for the 64$^3$ resolution, we find much
better agreement with the original gas run when following only the stars
making up the galaxy at z=0 (all star particles in a radius of 50 kpc). This
is shown in the thin green dash-dotted line. Applying the same procedure to
the 128$^3$ resolution results in the lower, blue dash-triple-dotted
line. The discontinuity to the gas run at z=1.7 is due to the switch in
calculation methods. The resulting angular momentum evolution of the two
calculation methods are not very different. The result of the real gas disk is
expected to be in between the two. In both cases, there is good agreement with
the evolution in 
the lower resolution runs, strongly indicating a gravitational effect as the
reason for the angular momentum loss and ruling out artefacts from the
hydrodynamics or SPH implementation in the code. The behavior in
all four resolutions is quite similar. For the highest resolution the loss is
slightly stronger, indicating that resolution makes the problem more severe
instead of improving it. We first calculate the ratio
of the final over the initial specific angular momentum of the disk particles
for our four resolutions. Over our entire resolution
range we find the same net loss of angular
momentum of about 55\% and no trend with resolution is observed. This indicates that
a simple increase in resolution does not help to alleviate the angular momentum
problem within the framework of the basic hierarchical model of galaxy
formation.\par
\begin{figure}
\begin{center}
\includegraphics[width=0.5\textwidth]{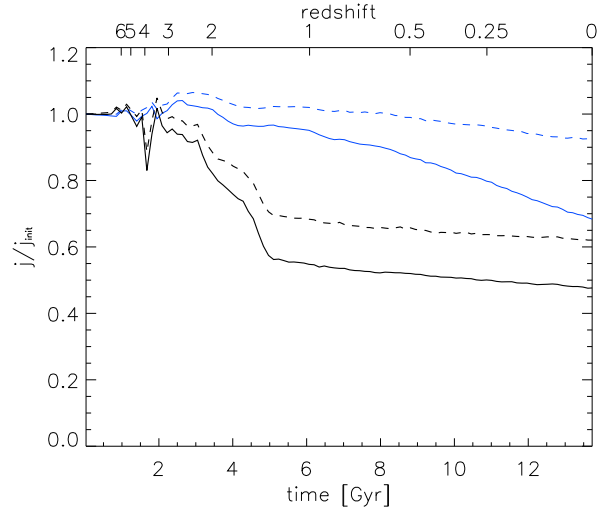}
\caption{Comparing the angular momentum evolution for the 32$^3$ halo in the
  standard case (with radiative cooling switched on all the time, black thick curves) with the run with adiabatic
  evolution until z=1 and cooling after z=1 (blue thin curves). The solid lines are
  for disk gas, the dashed lines for all gas within the virial radius at z=0.}
\label{fig:AMadiabatic}
\end{center}
\end{figure}

We now investigate, when and how the angular momentum is lost. In combination with
Figure~\ref{fig:semidiskevol}, Figure~\ref{fig:AMcomp} clearly illustrates
that the bulk of angular momentum is
lost during the phase of evolution dominated by the infall and merging of
smaller clumps. At z=1, the angular momentum ratio is j$_{\rm{z}=1}$/j$_{\rm{init}}\sim0.55$. Once the
disk is established at z$\sim$1, there is still some loss,
but this happens gradually with j$_{\rm{final}}$/j$_{\rm{z}=1}\sim0.88$. This further
illustrates that improved resolution does not strongly influence
this process, since the loss not only follows the same pattern
in all resolutions, it also would otherwise continue stronger even after the
formation of the disk.
We can further strengthen our conclusion by following \citet{Weil98} and
evolving the gas component adiabatically until z=1. The lack of radiative
cooling prevents the formation and therefore merging of small clumps in the
gas. The disk forms quickly after the cooling is turned on at
z=1. Figure~\ref{fig:AMadiabatic} shows the resulting difference for the gas
in the 32$^3$ run. There is still a loss of angular momentum, 
but it is smaller than before, with a final ratio for the disk of
j$_{\rm{final}}$/j$_{\rm{init}}$=68.3\% and j$_{\rm{z}=1}$/j$_{\rm{init}}$=95.1\% at z=1.

\subsection{Cosmological initial conditions}
\begin{figure}
\begin{center}
\includegraphics[width=0.5\textwidth]{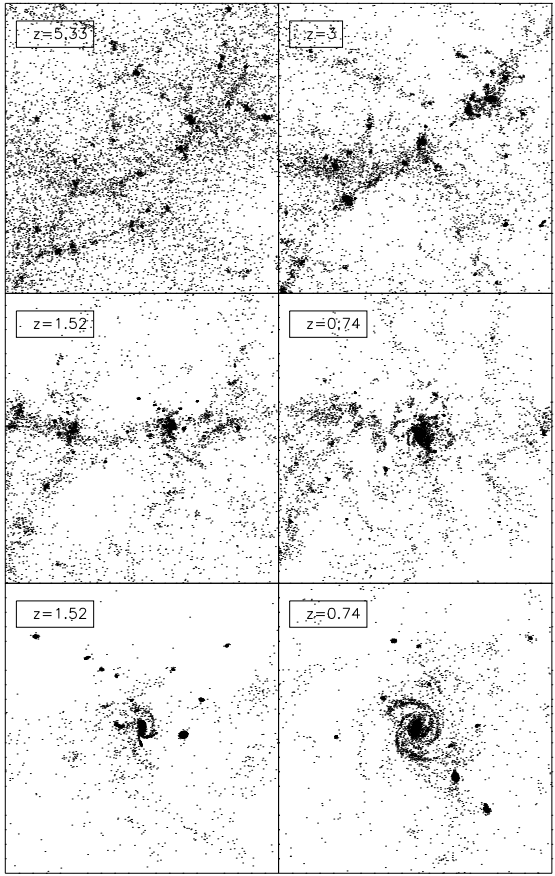}
\caption{Snapshots illustrating the assembly of the gas disk in the fully
  cosmological runs. The top four plots have a side length of 1000 comoving kpc
  h$^{-1}$ to show the larger environment, the bottom two plots zoom in to
  show more clearly the disk itself and has a side length of 200 kpc h$^{-1}$ in
  physical coordinates.}
\label{fig:builddisksnaps}
\end{center}
\end{figure}

\begin{figure}
\begin{center}
\includegraphics[width=0.5\textwidth]{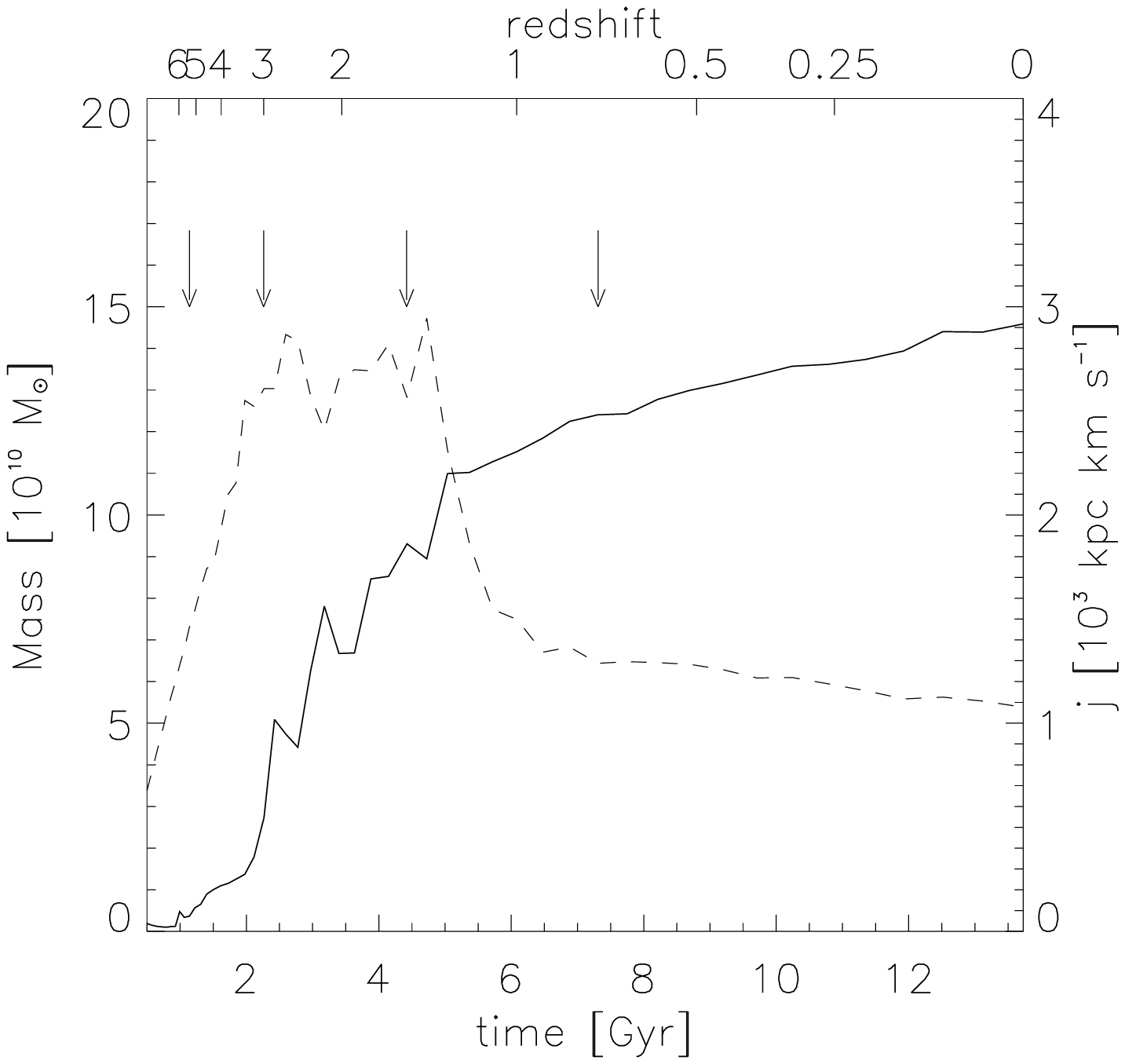}
\caption{Growth of the disk (solid line) and evolution of angular momentum
  (dashed line) for the fully cosmological run. The big loss in angular
  momentum starting at z=1.5 clearly coincides with a large mass increase. The
  arrows indicate the time of the snapshots shown in Figure~\ref{fig:builddisksnaps}.}
\label{fig:builddisk}
\end{center}
\end{figure}
We now turn our attention to the angular momentum problem in a fully
cosmological run to verify our conclusions. For this we use the
medium resolution cosmological halo, with 1024$^3$ particles effective
resolution in the high resolution region. Characteristic parameters of the
halo and disk can be found in Table~\ref{tab:stats}.\par
The disk in this case is as centrally concentrated as in the semi-cosmological
case with 78\% of the disk gas within the central 5 kpc. The build-up of the
gas disk is shown in Figure~\ref{fig:builddisksnaps}. The top and middle panels have
a side length of 1 h$^{-1}$ Mpc in comoving coordinates and nicely show the hierarchical
growth from small clumps to larger objects. In the time between the two middle
panels (z$\sim$1.5 and z$\sim$0.75) a last
larger merger of two gas clumps occurs. After that, the disk grows only via
accretion. The bottom panel zooms into the disk before and after that
merger. It is interesting to now compare (see Figure~\ref{fig:builddisk}) the
mass growth history of the disk (solid line) and the evolution of the specific
angular momentum of the disk particles (dashed line). The arrows indicate the
points in time at which the snapshots in Figure~\ref{fig:builddisksnaps} are
shown. The early growth of angular momentum accompanying the first collapse of
the halo follows well the expectations from
tidal torque theory (J$\propto$t). The maximum angular momentum acquired by the gas agrees
well with the angular momentum of the dark matter halo. However, it can clearly be
seen, that in particular the merger of two large gas clumps at z$\approx$1.5,
which increases the mass of the disk by about 25\%, is accompanied by a large
loss in angular momentum (about 50\%). The ratio of final to maximum specific
angular momentum is 0.37. This is a little less than what we found in the
semi-cosmological case, as expected, thus confirming and even strengthening our
conclusions of the previous section.

\section{The influence of star formation and stellar feedback on the angular momentum problem}
\label{sec:AM_feedback}
\begin{figure*}
\includegraphics[width=\textwidth]{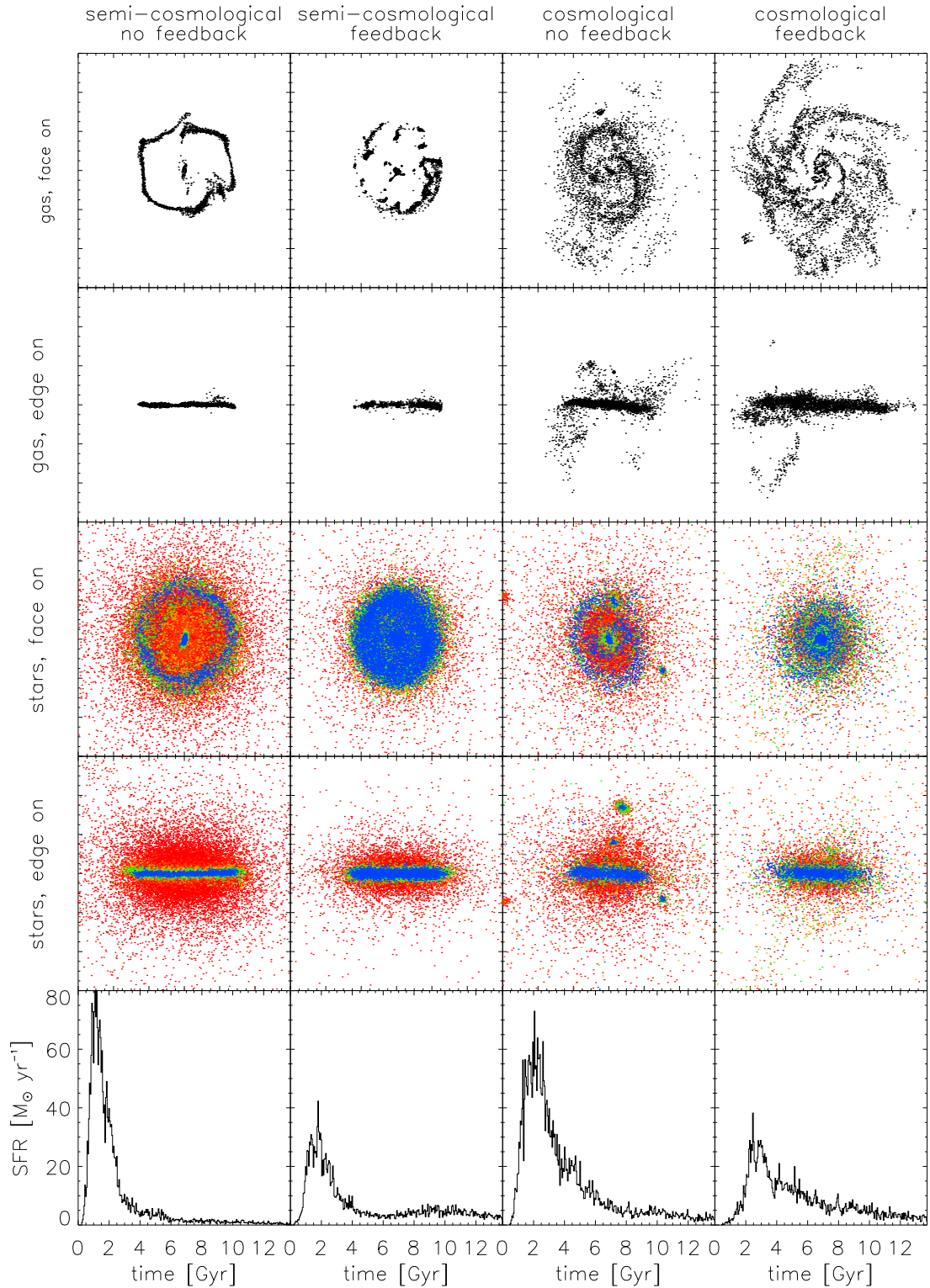}
\caption{Results at z=0 of the runs with star formation. In the first and third columns we show runs without feedback, and runs with feedback are displayed in the second and fourth column. The
  left two columns are semi-cosmological results, the right two columns are
  fully cosmological runs. The face on and edge on plots have a side length of 80
  kpc. The top two rows show the cold gaseous disk, rows 3 and 4 the stellar
  component. The star particles are color-coded by age (red: older
  than 10 Gyr, blue: younger than 4 Gyr). The bottom row shows the star
  formation history.}
\label{fig:SFresults}
\end{figure*}
In Figure~\ref{fig:SFresults}, we show the gas and stellar component of the
resulting galaxy in our simulations including star formation (columns 1 and 3
for the semi- and the cosmological environments, respectively) and including
star formation and feedback (columns 2 and 4
for the semi- and the cosmological environments, respectively). The first feature to
notice is the ring-like structure of the gas disk in the semi-cosmological
runs (top row, columns 1 and 2). The reason for this is the efficient transformation of gas into stars
moving inside-out, 
and the missing late infall of gas to replenish the disk. The disk here is
also much more defined than in the fully cosmological case, again due to the
missing influence of late incoming clumps. In all cases, the young stars (younger than 4 Gyr, colored in blue)
and the cold gas ($\rm{T}\leq3\times10^4$~K) form a fairly thin disk. In both the semi-cosmological
and the fully cosmological run without feedback (column 1 and 3) stars are
formed very efficiently at early times, resulting in a dominant spheroidal
component. This can also be seen in the strong peak in the star formation history at a t$\simeq$2 Gyr (bottom row). When feedback is
included (column 2 and 4), star formation is quenched efficiently and a more
prominent disk of young stars is formed while the large spheroidal component
made up of old stars (colored red) is reduced. It is also noticeable, that the
gas disk is much more extended than the young stellar component in the cosmological
runs. This is in agreement with observations \citep[for example][]{Boomsma2008}.
\subsection{The origin of the stellar components}
\begin{figure}
\begin{center}
\includegraphics[width=0.5\textwidth]{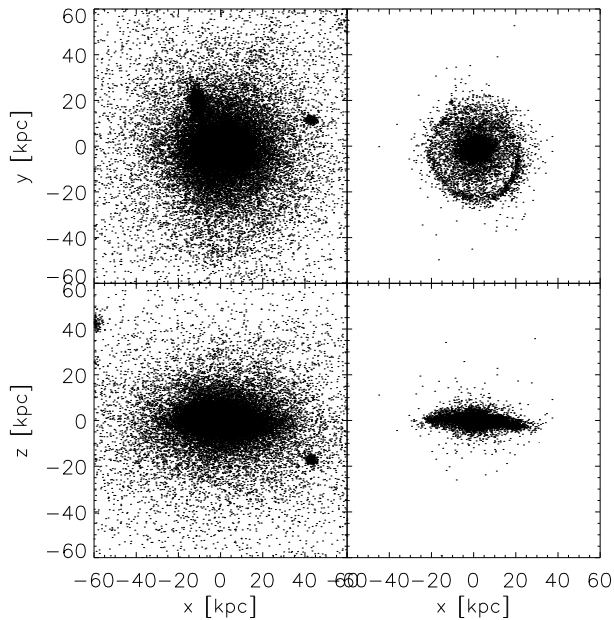}
\caption{Face-on and edge-on projection of the stars at redshift 0. The left
  column shows the stars which were formed in clumps and then accreted. The
  right column shows the stars which were formed in the disk progenitor.}
\label{fig:clumpdiskstars}
\end{center}
\end{figure}

\begin{figure}
\begin{center}
\includegraphics[width=0.5\textwidth]{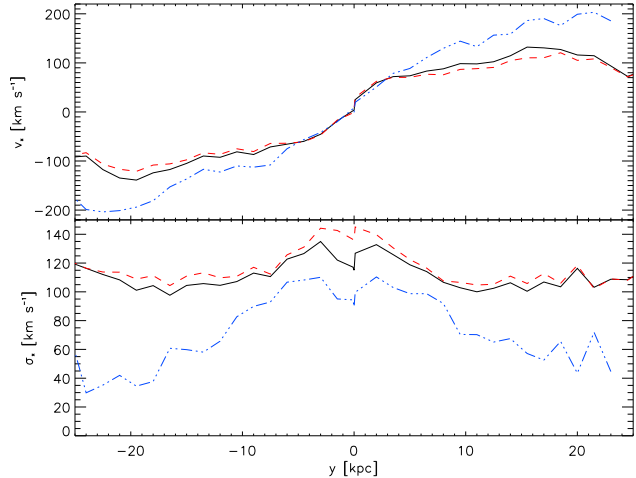}
\caption{Rotational
  velocity (top) and velocity dispersion (bottom) of all stars (black,solid
  line),"clump stars" (red, dashed line) and "disk stars" (blue, dash-triple-dotted line).}
\label{fig:veldisp}
\end{center}
\end{figure}

Why do we find such a large spheroidal component in the runs without feedback,
even though the gas disk in the runs without star formation is very thin? The
answer can be found in the location and the time where the stars are
formed. In our fully cosmological run, star formation starts at z$\sim$9, and it peaks
at z$\sim$3. This is the time when the progenitor of the z=0 disk is just
starting to be assembled, so the majority of the stars are actually forming in
larger mass clumps which fall in until z$\sim$1.2. After that, mass grows more
slowly through accretion and the infall of smaller clumps. Stars are
continuously formed in the disk progenitor, but at a very low rate of about 2
$\Msun\,\rm{yr}^{-1}$. The stars entering the halo in clumps are not able to
settle into a disk and therefore will form a spheroid. Only the stars formed within the disk progenitor
will be rotationally supported and can form a stellar disk. This can be seen
in Figure~\ref{fig:clumpdiskstars}, which shows the distribution of stars at
z=0, divided according to their formation environment into ``clump
stars'' (stars formed outside of the disk
progenitors and assembled into the galaxy via merging) and ``disk stars''
(stars formed in the progenitor of the z=0 disk). ``Disk stars'' form a reasonably thin distribution
tracing the gas disk, ``clump stars'' make up the large spheroid. One
important diagnostic of the nature of the stars is the ratio of rotational
velocity to velocity dispersion. Both parameters are plotted in
Figure~\ref{fig:veldisp} for all stars (black line), "clump stars" (red line)
and "disk stars" (blue line). For "disk stars", this ratio is much higher,
confirming their rotational support compared to dissipational support for
"clump stars". Of all stars
in the galaxy at z=0, 30\% are ``disk stars'' and 70\% are ``clump stars''. We
performed a test where star formation was only turned on at z=1, with the thin
gas disk being more or less in place and further mass infall mostly
through smooth accretion. The stars in this case indeed form a very prominent 
stellar disk, but no large spheroidal component. \par
In the feedback runs, star formation also begins early in individual
clumps. The feedback however heats the gas and pushes it out of these structures, quenching the
star formation in the clumps and enabling the gas do be accreted smoothly
later directly into the disk where it then can form disk stars. This is evident from the reduced early star formation peak in the feedback runs as shown in Figure~\ref{fig:SFresults}. Consequently,
in the final galaxy we have only 35\% of all stars formed in clumps (and
therefore a much less prominent spheroidal component as visible in
Figure~\ref{fig:SFresults} right column) and 64\%
formed in the disk progenitor.

\subsection{The angular momentum problem}
\begin{figure*}
\centering
\subfigure[]
{
  \includegraphics[width=0.45\textwidth]{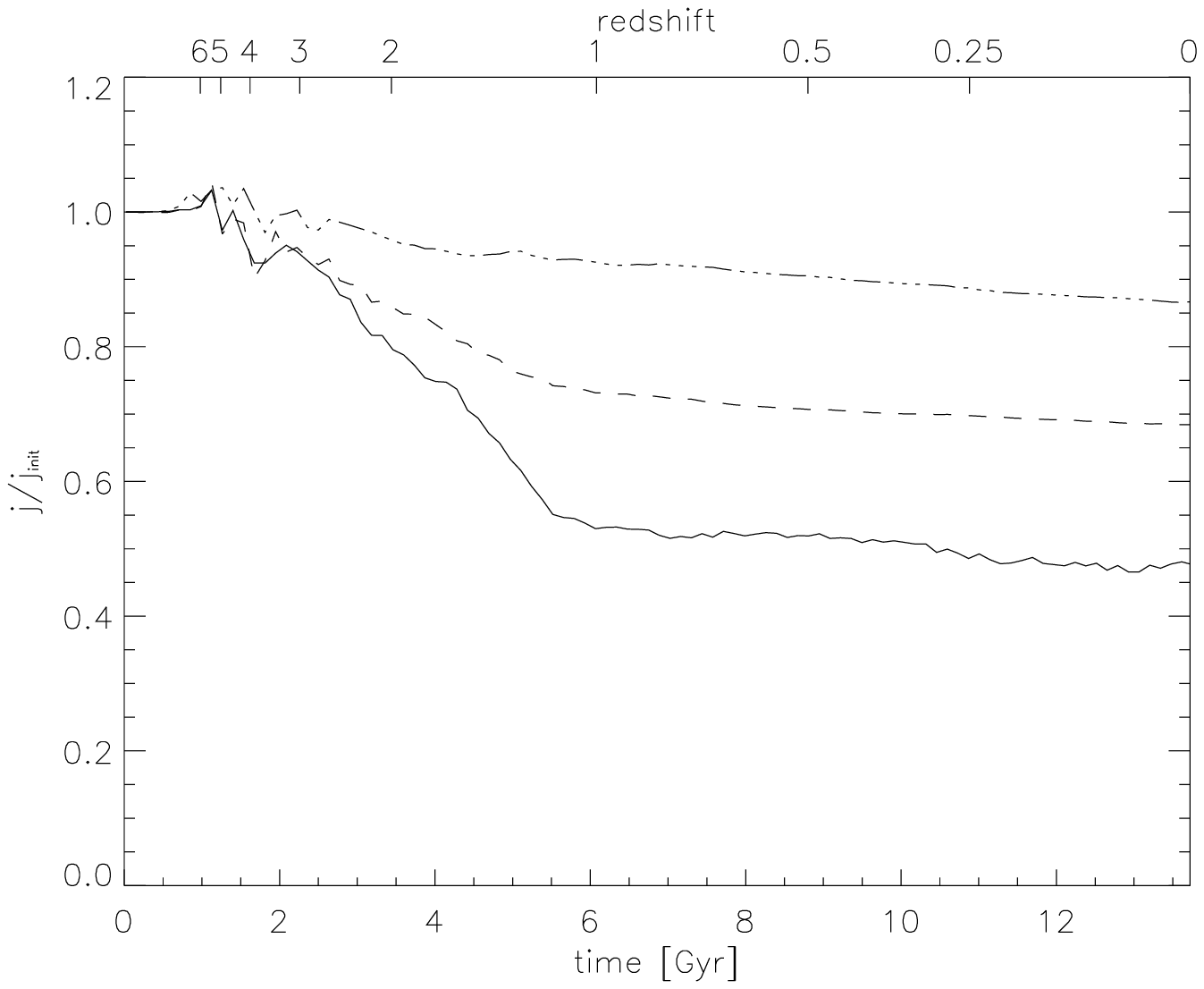}
  \label{fig:amloss_SFa}
}
\hspace{.5cm}
\subfigure[]
{
  \label{fig:amloss_SFb}
  \includegraphics[width=0.45\textwidth]{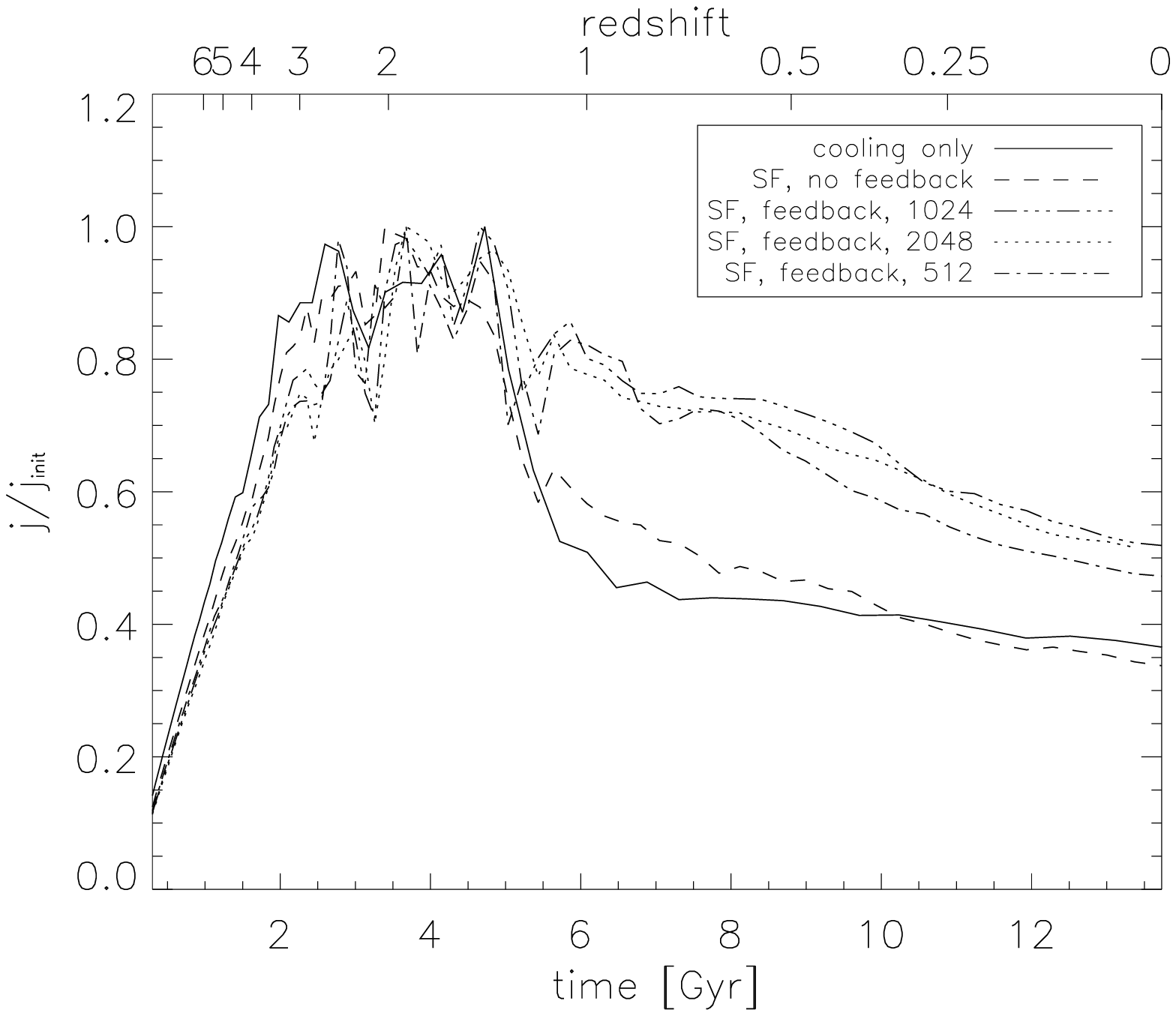}
}
\caption{These plots compare the specific angular momentum evolution in the
  runs without star formation (solid line), with star formation but without
  feedback (dashed line) and with star formation and feedback (dashed-dotted
  line). Panel (a) shows the semi-cosmological, panel (b) the fully
  cosmological runs. The three different lines for feedback in Panel (b) show
  runs with three different resolutions.} 
\label{fig:amloss_SFsub}
\end{figure*}
Since now gas particles are converted into stars, we compute the angular
momentum for the combination of cold gas and stars. The result for the
semi-cosmological runs is shown in Figure~\ref{fig:amloss_SFa}. The loss is smaller when including star
formation (dashed line), but still substantial. In the case with feedback
(dotted-dashed line), later accretion of gas with high angular momentum keeps the angular momentum of the disk high and the final galaxy has almost 90\% of the
initial angular momentum. Gas is available for late accretion due to the decreased early star formation. The results for the cosmological runs are shown in Figure~\ref{fig:amloss_SFb}. We are tracking the cold gas and stars of the
final galaxy at z=0 over time. In all three cases we see a drop in angular momentum at
z$\sim$1.2, but it is less pronounced in runs with star formation and even
smaller in those with
feedback. However, in simulations with star formation there is some loss of angular momentum 
after z=1, which is not as significant when only cooling is employed. This is probably due to the increased late accretion of gas and related hydrodynamical torques. Compared to
the maximum angular momentum, an even slightly higher percentage is lost in
the case with star formation, and still about 50\% with feedback. Most of this
loss occurs after z=1. The feedback has reduced the loss owing to dynamical friction
but increased the later loss due to hydrodynamical torques. While a loss of
50\% still seems large, this is consistent with expectations comparing dark
matter halos and observed disks \citep{NavarroMS2000}. For the 
feedback case, we plot the results of runs with three different resolutions
(effective resolutions of 512$^3$, 1024$^3$ and 2048$^3$), which behave  
similarly indicating that resolution does not affect this process much. We conclude that feedback in both the semi-cosmological and the cosmological  
simulations clearly helps to alleviate the angular momentum problem which is mainly
due to the reduction of the early star formation and the resulting smaller spheroidal component.

\section{Summary and Conclusions}
In this paper we have studied the causes of the angular momentum problem
with special emphasis on the influence of numerical resolution and the impact
of star formation and feedback. We performed a systematic study leading us 
from a controlled, isolated, "semi-cosmological" experiment to fully
cosmological simulations. We also investigated simulations with different
physical processes, once only taking
into account radiative cooling, then including simple star formation and finally also
including supernova feedback. The advantage of the semi-cosmological environment is to
combine at moderate computational costs an isolated, well controlled halo with the small scale hierarchical
buildup which is characteristic for a $\Lambda$CDM universe. By performing first simulations 
without star formation, 
it is possible to assess the importance of resolution in the
hierarchical framework for structure formation, especially with respect to the
angular momentum problem. Our results clearly show that an increase in
resolution in this case does not help against the angular momentum loss. The
loss of angular momentum for the disk gas is of the same order while the
resolution changes by orders of magnitude. We are also able to tie the angular
momentum loss to the phase of collapse into small cold gas clumps which
merge and form the main object. The main loss mechanism is dynamical friction
of the clumps and larger mergers, as shown by
\citet{NavarroBenz1991}, \citet{NavarroWhite1994}, \citet{navarro97},
\citet{Donghia} and \citet{Zavala08}. The efficiency of dynamical friction is
dominated by the largest clumps. These clumps can be tracked even in low
resolution simulations \citep[][and references therein]{Gao2004}. This explains why, contrary to for example
\citet{Kaufmann2007}, we find no significant resolution dependence even though our resolution range is
comparable. \citet{Kaufmann2007} studied the formation of a gas disk
starting in an equilibrium halo without taking the halo formation history into
account, so their angular momentum loss is mainly due to hydrodynamical and
gravitational torques during gas infall. This is a resolution-dependent
process. We also see this effect, as can be seen in the run with cooling
switched on only after z=1 (Figure~\ref{fig:AMadiabatic}) and in the
cosmological runs with star formation and feedback. However, the effect is
much smaller than the loss during the merger period. We therefore argue that,
in order to prevent the angular momentum loss in cosmological simulations,
mainly the early formation of gas and star  clumps must be prevented. The loss
owing to limited numerical resolution appears to be only of second order.
\begin{figure}
\begin{center}
\includegraphics[width=0.5\textwidth]{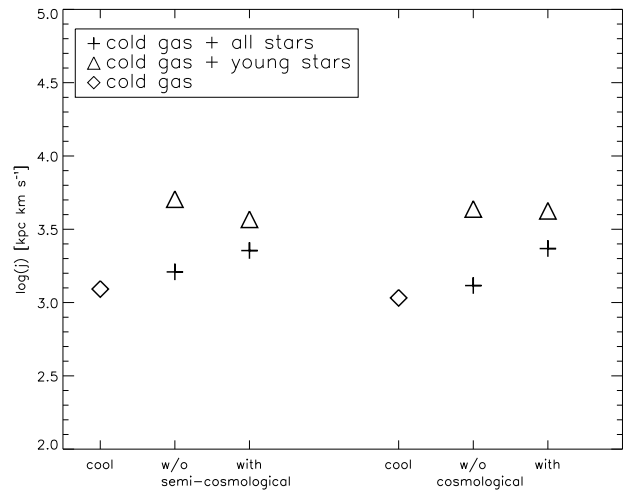}
\caption{For the semi-cosmological and cosmological run, we compare the final
  angular momentum of the disk in the cooling run (``cool''), with star formation but
  without feedback (``w/o'') and with feedback (``with''). The disk angular
  momentum is computed once
  including the cold gas disk and all stars in the galaxy (crosses), and
  also including the cold gas and the young stars only (triangles).}
\label{fig:amloss_starage}
\end{center}
\end{figure}
In line with this argument, we observe a comparable loss in our fully
cosmological simulation. It can be clearly linked again to the merging of two
large gas clumps. Unfortunately it is computationally not feasible to run this
at higher resolution without star formation. However, even if resolution would
have some influence in cosmological simulations (contrary to the
semi-cosmological case), this would only make the problem worse by increasing
the number of resolved gas clumps. We are confident that this would not lead
us to a different conclusion.\par 
The most plausible mechanism to prevent
clump formation is star formation and related supernova feedback because it
heats the gas. Another possible source of heating would be the cosmic UV
background. Some low resolution tests we performed including this effect did
not show any positive impact. Indeed, \citet{navarro97} showed that a UV
background may even have an adverse effect as it suppresses the infall of
gas which is accreted late in the formation history and thus acquired high angular
momentum. Here we show the changes in the angular momentum evolution when star
formation and supernova feedback is included. As expected, star
formation alone does not result in an improvement, since the same clumps are
spiraling in. They simply consist of highly condensed stars instead of highly
condensed gas. Star formation
peaks high early on, and especially in the semi-cosmological setup, where late
gas infall is missing, most gas is used up at z=0. Feedback helps to prevent
this by keeping the gas too hot to form stars and therefore allowing it to
settle into the disk later on as high angular momentum gas. We studied the origin
of the stars in the final galaxy and found that without feedback 70\% of all
stars in the final galaxy enter in
clumps. With feedback this is reduced to 35\%. This can also be seen in the
angular momentum of the final galaxy. Clump stars form a slowly rotating
spheroidal component and disk stars form a rotationally supported, high
angular momentum disk. The final angular momentum of the gas and all stars
combined is therefore much higher in our runs with feedback (and a smaller
spheroid) than in the ones without feedback (and a large spheroid). This is
further illustrated when looking at the combined angular momentum of the
rotationally supported component of the galaxy, consisting of the cold gas and
the young stars only (Figure~\ref{fig:amloss_starage}). This is roughly equal
in the runs with and without feedback (triangles), while for the total
galactic angular momentum including all stars a clear increase can be seen
from the run including only radiative cooling over the run with star formation
to the run with star formation and feedback. This increase is due to the
decrease of the central gas clump and then of the stellar spheroid.\\%\par
We conclude that an efficient suppression of the formation of dense clumps at
early times is the main factor in reducing the angular momentum loss. Star
formation alone cannot provide this effect since these dense gas clumps also
fulfill the conditions for star formation resulting in a high early star
formation peak and the formation of a large bulge. Only feedback can reduce
this effect efficiently. Resolution overall plays a secondary role. In our simulations with
star formation and feedback, the angular momentum loss is about 50\% which is in agreement with
expectations from comparisons between dark matter halos and observed disk
galaxies \citep{NavarroMS2000}. This loss mainly happens at late times due to
hydrodynamical torques. In our simulations with three levels of resolution in
the cosmological halo, resolution does not influence this
effect either. While feedback has such a critical role in
the formation of more realistic disks in simulations, it is still not well
understood. In \citet{Paper2} we therefore present a detailed study of stellar
feedback in disk formation simulations.
\section*{Acknowledgements}
We thank Stefan Gottl\"ober and Gustavo Yepes for providing initial conditions
and Volker Springel for providing a version of \textsc{GADGET2} which included
radiative cooling. This work was funded through a grant by the German Research
Foundation (DFG) under STE 710/4 as part of the Priority Programme SPP1177
"Witnesses of Cosmic History: Formation and evolution of black holes, galaxies
and their environment".

\label{lastpage}


\begin{thebibliography}{}
\bibitem[Barnes \& White(1984)]{BW84} Barnes, J., White, S.D.M., 1984, MNRAS,
  211, 753
\bibitem[Barnes \& Efstathiou(1987)]{BE87} Barnes, J, Efstathiou, G., 1987,
  ApJ, 319, 575
\bibitem[Blumenthal et al.(1986)]{Blum86} Blumenthal, G.R., Faber, S.M.,
  Flores, R., Primack, J.R., 1986, ApJ, 301, 27
\bibitem[Boomsma et al.(2008)]{Boomsma2008} Boomsma, R., Oosterloo, T.A.,
  Fraternali, F., van der Hulst, J.M., Sancisi, R., 2008, A\&A, 490, 555
\bibitem[Bullock et al.(2001)]{bullock01} Bullock, J.S., Dekel, A., Kolatt,
  T.S., Kravtsov, A.V., Klypin, A.A., Porciani, C., Primack, J.R., 2001, ApJ,
  555, 240
\bibitem[Doroshkevich(1970)]{Doroshkevich70} Doroshkevich, A.G., 1970,
  Astrofisika, 6, 581
\bibitem[d'Onghia et al.(2006)]{Donghia} D'Onghia, E., Burkert, A., Murante,
  G., Khochfar, S., 2006, MNRAS, 372, 1525
\bibitem[Duerr et al.(1982)]{Duerr1982} Duerr, R., Imhoff, C.L., Lada, C.J.,
  1982, ApJ, 261, 135
\bibitem[Fall \& Efstathiou(1980)]{FallEfs1980} Fall, S.M., Efstathiou, G., 1980,
  MNRAS, 193,189
\bibitem[Gao et al.(2004)]{Gao2004} Gao, L., White, S.D.M., Jenkins, A.,
  Stoehr, F., Springel, V., 2004, MNRAS, 355, 819
\bibitem[Gerritsen(1997)]{Gerritsen1997} Gerritsen, J.P.E., 1997, PhD thesis,
  Kapteyn Astron.Inst., The Netherlands
\bibitem[Gingold \& Monaghan(1977)]{gm77} Gingold, R.A., Monaghan, J.J.,
  1977, MNRAS, 181, 375
\bibitem[G\"otz \& Sommer-Larsen(2003)]{Goetz2003} G\"otz, M., Sommer-Larsen,
  J., 2003, Ap\&SS, 284, 341
\bibitem[Gottbrath(2001)]{Gottbrath} Gottbrath, C., 2001, Master's thesis,
  University of Arizona, USA
\bibitem[Governato et al.(2004)]{fabio2004} Governato, F., Mayer, L., Wadsley,
  J., Gardner, J.P., Willman, B., Hayashi, E., Quinn, T., Stadel, J., Lake,
  G., 2004, ApJ, 607, 688
\bibitem[Hoyle(1949)]{hoyle} Hoyle, F., 1949, in \emph{Problems of Cosmical
    Aerodynamics} (Dayton, Ohio: Central Air Documents Office), p. 195
\bibitem[Katz(1991)]{Katz1991} Katz, N., 1991, ApJ, 368, 325
\bibitem[Katz(1992)]{Katz1992} Katz, N., 1992, ApJ, 391, 502
\bibitem[Katz et al.(1996)]{Katz96} Katz, N., Weinberg, D.H., Hernquist, L.,
  1996, ApJS, 105, 19
\bibitem[Kaufmann et al.(2007)]{Kaufmann2007} Kaufmann, T., Mayer, L., Wadsley, J.,
  Stadel, J., Moore, B., 2007, MNRAS, 375, 53
\bibitem[Kautsch et al.(2006)]{kautsch} Kautsch, S.J., Grebel, E.K., Barazza,
  F.D., Gallagher, J.S.,III, 2006, A\&A, 445, 765
\bibitem[Lokas \& Hoffman(2001)]{Lokas2001} Lokas, E.L., Hoffman, Y., 2001, 
  arXiv:astro-ph/0112031 
\bibitem[Lucy(1977)]{lucy77} Lucy, L.B., 1977, ApJ, 82, 1013
\bibitem[Mayer(2005)]{Mayer2005} Mayer, L., 2005, Proceedings of Science,
  volume of the Conference ``Baryons in Dark Matter halos'', edited by
  P. Salucci, Novigrad, Croatia
\bibitem[Miller \& Scalo(1979)]{millerscalo} Miller, G.E., Scalo, J.M., 1979,
  ApJS, 41, 51
\bibitem[Mo, Mao \& White(1998)]{MMW1998} Mo, H.J., Mao, S., White, S.D.M., 1998,
  MNRAS, 295, 319
\bibitem[Navarro \& Benz(1991)]{NavarroBenz1991} Navarro, J.F., Benz, W., 1991, ApJ,
  380, 320
\bibitem[Navarro \& White(1994)]{NavarroWhite1994} Navarro, J.F., White, S.D.M., 1994,
  MNRAS, 267, 401
\bibitem[Navarro, Frenk \& White(1997)]{Navarro1997} Navarro, J.F., Frenk, C.S.,
  White, S.D.M., 1997, ApJ, 490, 493
\bibitem[Navarro \& Steinmetz(1997)]{navarro97} Navarro, J.F., Steinmetz, M.,
  1997, ApJ, 478, 13
\bibitem[Navarro \& Steinmetz(2000)]{NavarroMS2000} Navarro, J.F., Steinmetz, M.,
  2000, ApJ, 538, 447
\bibitem[Okamoto et al.(2005)]{okamoto03} Okamoto, T., Eke, V.r., Frenck, C.S.,
  Jenkins, A., 2005, MNRAS, 363, 1299
\bibitem[Peebles(1969)]{Peebles69} Peebles, P.J.E., 1969, ApJ, 155, 393
\bibitem[Piontek \& Steinmetz(2009b)]{Paper2} Piontek, F., Steinmetz, M., 2009,
  submitted to MNRAS
\bibitem[Power et al.(2003)]{Power2003} Power, C., Navarro, J.F., Jenkins, A.,
  Frenk, C.S., White, S.D.M., Springel, V., Stadel, J., Quinn, T., 2003,
  MNRAS, 338, 14
\bibitem[Robertson et al.(2004)]{Robertson2004} Robertson, B., Yoshida, N., Springel,
  V., Hernquist, L., 2004, ApJ, 606, 32
\bibitem[Schmidt(1959)]{schmidt59} Schmidt, M., 1995, ApJ, 129, 243
\bibitem[Sommer-Larsen \& Dolgov(2001)]{SommerLarsen2001} Sommer-Larsen, J., Dolgov, A.,
  2001, ApJ, 551, 608
\bibitem[Spergel et al.(2007)]{WMAP3year} Spergel, D.N. et al., 2007, ApJS,
  170, 377
\bibitem[Springel(2005)]{Springel2005} Springel, V., 2005, MNRAS, 364, 1105
\bibitem[Steinmetz \& Bartelmann(1995)]{MS95} Steinmetz, M., Bartelmann, M.,
  1995, MNRAS, 272, 570
\bibitem[Steinmetz \& M\"uller(1995)]{MSMueller95} Steinmetz, M., M\"uller,
  E., 1995, MNRAS, 276, 549
\bibitem[Steinmetz \& Navarro(1999)]{MSNavarro1999} Steinmetz, M., Navarro,
  J.F., 1999, ApJ, 513, 555
\bibitem[Steinmetz \& White(1997)]{MSWhite1997} Steinmetz, M., White, S.D.M.,
  1997, MNRAS, 288, 545
\bibitem[Str\"omberg(1934)]{stromberg} Str\"omberg, G., 1934, ApJ, 79, 460
\bibitem[Thacker \& Couchman(2000)]{thack00} Thacker, R.J., Couchman, H.M.P.,
  2000, ApJ, 545, 728
\bibitem[Toomre(1964)]{toomre} Toomre, A., 1964, ApJ, 139, 1217
\bibitem[Toomre \& Toomre(1972)]{toomre72} Toomre, A., Toomre, J., 1972, ApJ,
  178, 623
\bibitem[Viel et al.(2008)]{Viel2008} Viel, M., Becker, G.D., Bolton, J.S.,
  Haehnelt, M.G.,. Rauch, M., Sargent, W.L.W., 2008, PhRvL, 100, 041304
\bibitem[Weil, Eke \& Efstathiou(1998)]{Weil98} Weil, M.L., Eke, V.R.,
  Efstathiou, G., 1998, MNRAS, 300, 773
\bibitem[White \& Rees(1978)]{white78} White, S.D.M., Rees, M.J., 1978,
  MNRAS, 183, 341
\bibitem[White(1984)]{white84} White, S.D.M., 1984, ApJ, 286, 38
\bibitem[Zavala et al.(2008)]{Zavala08} Zavala, J., Okamoto, T., Frenk, C.S.,
  2008, MNRAS, 387, 364
\end{thebibliography}
\end{document}